\documentclass[aps,prl,twocolumn,tightenlines,superscriptaddress,showpacs,byrevtex]{revtex4}

\usepackage{graphicx}
\usepackage{dcolumn}
\usepackage{bm}
\begin{document}
 
\title{Measurement of the branching fractions for 
$B^- \to D^{(*)+} \pi^- \ell^{-} \bar{\nu}_{\ell}$ and 
$\bar{B^0} \to D^{(*)0} \pi^+ \ell^{-} \bar{\nu}_{\ell}$}

\affiliation{Budker Institute of Nuclear Physics, Novosibirsk}
\affiliation{Chonnam National University, Kwangju}
\affiliation{University of Cincinnati, Cincinnati, Ohio 45221}
\affiliation{University of Hawaii, Honolulu, Hawaii 96822}
\affiliation{High Energy Accelerator Research Organization (KEK), Tsukuba}
\affiliation{Institute of High Energy Physics, Chinese Academy of Sciences, Beijing}
\affiliation{Institute of High Energy Physics, Vienna}
\affiliation{Institute for Theoretical and Experimental Physics, Moscow}
\affiliation{J. Stefan Institute, Ljubljana}
\affiliation{Kanagawa University, Yokohama}
\affiliation{Korea University, Seoul}
\affiliation{Swiss Federal Institute of Technology of Lausanne, EPFL, Lausanne}
\affiliation{University of Ljubljana, Ljubljana}
\affiliation{University of Maribor, Maribor}
\affiliation{University of Melbourne, Victoria}
\affiliation{Nagoya University, Nagoya}
\affiliation{Nara Women's University, Nara}
\affiliation{National Central University, Chung-li}
\affiliation{National United University, Miao Li}
\affiliation{Department of Physics, National Taiwan University, Taipei}
\affiliation{H. Niewodniczanski Institute of Nuclear Physics, Krakow}
\affiliation{Nippon Dental University, Niigata}
\affiliation{Niigata University, Niigata}
\affiliation{Nova Gorica Polytechnic, Nova Gorica}
\affiliation{Osaka City University, Osaka}
\affiliation{Osaka University, Osaka}
\affiliation{Panjab University, Chandigarh}
\affiliation{Peking University, Beijing}
\affiliation{Princeton University, Princeton, New Jersey 08544}
\affiliation{University of Science and Technology of China, Hefei}
\affiliation{Shinshu University, Nagano}
\affiliation{Sungkyunkwan University, Suwon}
\affiliation{University of Sydney, Sydney NSW}
\affiliation{Tata Institute of Fundamental Research, Bombay}
\affiliation{Toho University, Funabashi}
\affiliation{Tohoku Gakuin University, Tagajo}
\affiliation{Tohoku University, Sendai}
\affiliation{Department of Physics, University of Tokyo, Tokyo}
\affiliation{Tokyo Institute of Technology, Tokyo}
\affiliation{Tokyo Metropolitan University, Tokyo}
\affiliation{Tokyo University of Agriculture and Technology, Tokyo}
\affiliation{University of Tsukuba, Tsukuba}
\affiliation{Virginia Polytechnic Institute and State University, Blacksburg, Virginia 24061}
\affiliation{Yonsei University, Seoul}

   \author{D.~Liventsev}\affiliation{Institute for Theoretical and Experimental Physics, Moscow} 
   \author{T.~Matsumoto}\affiliation{Tokyo Metropolitan University, Tokyo} 
   \author{K.~Abe}\affiliation{High Energy Accelerator Research Organization (KEK), Tsukuba} 
   \author{K.~Abe}\affiliation{Tohoku Gakuin University, Tagajo} 
   \author{H.~Aihara}\affiliation{Department of Physics, University of Tokyo, Tokyo} 
   \author{Y.~Asano}\affiliation{University of Tsukuba, Tsukuba} 
   \author{T.~Aushev}\affiliation{Institute for Theoretical and Experimental Physics, Moscow} 
   \author{A.~M.~Bakich}\affiliation{University of Sydney, Sydney NSW} 
   \author{V.~Balagura}\affiliation{Institute for Theoretical and Experimental Physics, Moscow} 
   \author{M.~Barbero}\affiliation{University of Hawaii, Honolulu, Hawaii 96822} 
% ...
\collaboration{The Belle Collaboration}

\date{\today}

\begin{abstract}
We report on a measurement of the branching fractions for 
$B^- \to D^{(*)+} \pi^- \ell^{-} \bar{\nu}_{\ell}$ and 
$\bar{B^0} \to D^{(*)0} \pi^+ \ell^{-} \bar{\nu}_{\ell}$ 
with 275~million $B\bar{B}$ events collected at the $\Upsilon(4S)$ 
resonance with the Belle detector at KEKB. 
Events are tagged by fully reconstructing one of the $B$ mesons
in hadronic modes. We obtain 
${\cal{B}}( B^- \to D^+ \pi^- \ell^- \bar{\nu}_{\ell} ) = (0.54 \pm 0.07({\rm stat}) \pm 0.07({\rm syst}) \pm 0.06({\rm BR}) ){\times}10^{-2}$,  
${\cal{B}}( B^- \to D^{*+} \pi^- \ell^- \bar{\nu}_{\ell} ) = (0.67 \pm 0.11({\rm stat}) \pm 0.09({\rm syst}) \pm 0.03({\rm BR}) ){\times}10^{-2}$, 
${\cal{B}}( \bar{B^0} \to D^0 \pi^+ \ell^- \bar{\nu}_{\ell} ) = (0.33 \pm 0.06({\rm stat}) \pm 0.06({\rm syst}) \pm 0.03({\rm BR}) ){\times}10^{-2}$, 
${\cal{B}}( \bar{B^0} \to D^{*0} \pi^+ \ell^- \bar{\nu}_{\ell} ) = (0.65 \pm 0.12({\rm stat}) \pm 0.08({\rm syst}) \pm 0.05({\rm BR}) ){\times}10^{-2}$, 
where the third error comes from the error on $\bar{B} \to D^{(*)}\ell^{-} \bar{\nu}_{\ell}$ decays. Contributions from $\bar{B^0} \to D^{*+} \ell^- \bar{\nu}_{\ell}$ decays
are excluded in the measurement of $\bar{B^0} \to D^0 \pi^+ \ell^- \bar{\nu}_{\ell}$.
\end{abstract}


\maketitle

Semileptonic decays play a prominent role in the study of $B$~meson properties. 
The total semileptonic branching fraction has been precisely 
determined to be $(10.73 \pm 0.28){\times}10^{-2}$~\cite{PDG2004}. 
While $\bar{B} \to D \ell^{-} \bar{\nu}_{\ell}$ and $D^{*} \ell^{-} \bar{\nu}_{\ell}$ account for $70 \%$ of this total, other contributions are not yet well understood. 
The most promising candidates include resonant and non-resonant $D^{(*)}\pi$ in the final state.
Both ALEPH~\cite{ALEPH_ddlnu} and DELPHI~\cite{DELPHI_ddlnu} have
studied $\bar{B} \to D^{(*)} \pi \ell^{-} \bar{\nu}_{\ell}$ 
decays (where both $\pi^{\pm}$ and $\pi^0$ modes are included).  
Assuming that semileptonic $B$ decays other than 
$\bar{B} \to D \ell^{-} \bar{\nu}_{\ell}$ and $\bar{B} \to D^{*} \ell^{-} \bar{\nu}_{\ell}$ 
are of the form $\bar{B} \to D^{(*)} \pi \ell^{-} \bar{\nu}_{\ell}$, they find:
\begin{eqnarray*}
   {\cal{B}}_{D^{(*)} \pi \ell^{-} \bar{\nu}_{\ell}}
   &\equiv& {\cal{B}}(\bar{B} \to D \pi \ell^{-} \bar{\nu}_{\ell}) + {\cal{B}}(\bar{B} \to D^{*} \pi \ell^{-} \bar{\nu}_{\ell})\\
   &=& ( 2.16 \pm 0.30 \pm 0.30 ){\times}10^{-2}~{\rm(ALEPH)},\\
   &=& ( 3.40 \pm 0.52 \pm 0.32 ){\times}10^{-2}~{\rm(DELPHI)}.
   \end{eqnarray*}
The former result suggests that there is a significant unknown contribution 
to the semileptonic branching fraction, while the latter shows no such deficit.
More precise measurements are therefore desired to resolve this discrepancy and clarify the difference between the rate for the inclusive semileptonic 
decay and the sum of the rates for the exclusive modes.
Improvement in the knowledge of the $\bar{B} \to D^{(*)} \pi \ell^{-} \bar{\nu}_{\ell}$ branching
fractions will also help to reduce systematic uncertainties in 
measurements of Cabibbo-Kobayashi-Maskawa elements such as 
$|V_{cb}|$ and $|V_{ub}|$~\cite{Belle_Vxb}.

In this paper, we present measurements of the branching fractions for
$\bar{B} \to D^{(*)} \pi \ell^{-} \bar{\nu}_{\ell}$ decays.
Inclusion of charge conjugate decays is implied throughout the paper. 
The analysis is based on data collected with the Belle detector~\cite{Belle} 
at the KEKB $e^+e^-$ asymmetric collider~\cite{KEKB}. 
We use a 253~fb$^{-1}$  data sample at the $\Upsilon(4S)$ 
resonance ($\sqrt{s} \simeq 10.58~{\rm GeV}$), 
corresponding to a sample of 275 million $B\bar{B}$ pairs. 
The selection of hadronic events is described elsewhere~\cite{Hadron_select}.
An additional 28 fb$^{-1}$ data sample taken 
at a center-of-mass energy 60 MeV below the $\Upsilon(4S)$ resonance is also used 
to study continuum $e^+e^- \to q\bar{q}\;(q=u,d,s,c)$ events.

The Belle detector is a large-solid-angle spectrometer with a 1.5 T magnetic
field provided by a super-conducting solenoid coil. Charged particles are measured using 
hits in a silicon vertex detector (SVD) and a 50-layer central drift chamber (CDC).
Photons are detected in an electromagnetic calorimeter (ECL) comprised of CsI(Tl) crystals.
Kaon identification is performed by combining the responses from an array of 
aerogel threshold \v{C}erenkov counters (ACC), a barrel-like arrangement 
of time-of-flight scintillation counters (TOF), and $dE/dx$ measurements in the CDC. 
A $K/\pi$ likelihood ratio $P_{K/\pi}$, ranging from 0 (likely to be a pion) to 
1 (likely to be a kaon), is formed. 
With the requirement $P_{K/\pi} > 0.6$, the kaon efficiency is approximately 
88\% and the average pion mis-identification rate is about 8\%. 
Electron identification is based on a combination of $dE/dx$ in CDC,
the response of ACC, shower shape in ECL, and the ratio of energy 
deposit in ECL to the momentum measured by the tracking system.
Muon identification is performed using resistive counters interleaved in the iron yoke,
located outside the coil. 
The lepton identification efficiencies are about 90\% 
for both electrons and muons in the momentum region above 1.2 GeV/$c$, 
where leptons from the prompt $B$ decays dominate.
The hadron mis-identification rate is less than 0.5\% for electrons and 
2\% for muons in the same momentum region.

We use a GEANT-based Monte Carlo (MC) simulation to model 
the response of the detector and determine its acceptance~\cite{GEANT}.
$\bar{B} \to D^{(*)} \ell^{-} \bar{\nu}_{\ell}$ and 
$D^{(*)} \pi \ell^{-} \bar{\nu}_{\ell}$ events are modeled using the EvtGen program~\cite{Evtgen}.
We use an HQET-based model~\cite{HQET2} for $\bar{B} \to D^{(*)} \ell^{-} \bar{\nu}_{\ell}$ 
and the Goity-Roberts model~\cite{Goity_Roberts} for $\bar{B} \to D^{(*)} \pi \ell^{-} \bar{\nu}_{\ell}$.
$\bar{B} \to D^{**} ( \to D^{(*)} \pi ) \ell^{-} \bar{\nu}_{\ell}$ is 
also simulated using the ISGW model~\cite{ISGW2} to evaluate the model 
dependence.

To suppress the high combinatorial background expected in the reconstruction 
of final states including a neutrino, we fully reconstruct one of the $B$ mesons, referred to hereafter as the {\it tag}. 
This allows us to separate particles created in the tag decay from 
those used in reconstructing the semileptonic decay, which we 
call {\it signal}.  It also provides a measurement
of the momentum of the signal $B$ meson, thus greatly improving the
resolution on the missing momentum.

The tag is fully reconstructed in the following modes:
$B^+ \to \bar{D}^{(*)0} \pi^+$, $\bar{D}^{(*)0} \rho^+$, 
$\bar{D}^{(*)0} a_1^+$, $\bar{D}^{(*)0}D_s^{(*)+}$, and 
$B^0 \to D^{(*)-} \pi^+$, $D^{(*)-} \rho^+$, $D^{(*)-} a_1^+$, 
$D^{(*)-} D_s^{(*)+}$. $\bar{D^0}$ candidates are reconstructed in seven 
modes: $\bar{D^0} \to K^+ \pi^-$, $K^+ \pi^- \pi^0$, $K^+ \pi^+ \pi^+ \pi^-$, 
$K_S^0 \pi^0$, $K_S^0 \pi^+ \pi^-$, $K_S^0 \pi^+ \pi^- \pi^0$, $K^+ K^-$. 
$D^-$ candidates are reconstructed in six modes: 
$D^- \to K^+ \pi^- \pi^-$, $K^+ \pi^+ \pi^- \pi^0$, $K_S^0 \pi^-$, 
$K_S^0 \pi^- \pi^0$, $K_S^0 \pi^+ \pi^- \pi^-$, $K^+ K^- \pi^-$. 
$D_s^+$ candidates are reconstructed in $D_s^+ \to K_S^0 K^+$, $K^+ K^- \pi^+$.
The $D$ candidates are required to have an invariant mass $m_D$ within $\pm 4 - 5 \sigma$ 
of the nominal $D$ mass value depending on the mode. 
$D^*$ mesons are reconstructed in the $D^{*+} \to D^0\pi^{+}/D^+\pi^0$, 
$D^{*0} \to D^{0}\pi^{0}/D^{0}\gamma$ and $D_{s}^{*+} \to D_s^{+} \gamma$ decays. 
$D^{*}$ candidates from modes that include a pion are required to have a mass 
difference $\Delta{m} = m_{D\pi} - m_{D}$ within $\pm 5~{\rm MeV}/c^2$ 
of its nominal value. For the decays with a photon, we require that the mass difference $\Delta{m} = m_{D\gamma} - m_{D}$ be within $\pm 20~{\rm MeV}/c^2$ 
of the nominal value.

The selection of the tag candidates is based on 
 $M_{\rm bc} = \sqrt{ E_{\rm beam}^2 - p_B^2 }$ and  $\Delta{E} = E_B - E_{\rm beam}$,
where $E_{\rm beam} \equiv \sqrt{s}/2 \simeq 5.29~{\rm GeV}$, and 
$p_B$ and $E_B$ are the momentum and  energy of the reconstructed $B$
in the $\Upsilon(4S)$ rest frame, respectively. The background from 
jet-like continuum events is suppressed on the basis of event
topology: we require the normalized second Fox-Wolfram moment~\cite{R2} 
to be smaller than $0.5$, and $|\cos\theta_{\rm th}|<0.8$, where $\theta_{\rm th}$ 
is the angle between the thrust axis of the $B$ candidate and that of 
the remaining tracks in an event. The latter requirement is not applied to 
$B^+\to\bar{D^0}\pi^+,\bar{D}^{*0}({\to}\bar{D}^0\pi^0)\pi^{+}$ and 
$B^0{\to}D^{*-}(\to\bar{D^0}\pi^-)\pi^+$ decays, where this background is smaller.

The signal region for tag candidates is defined 
as $M_{\rm bc} > 5.27~{\rm GeV}/c^2$ and $-80~{\rm MeV} < \Delta{E} < 60~{\rm MeV}$. 
If an event has multiple $B$ candidates, we choose
for each of $B^+$ and $B^0$ the candidate with the smallest 
$\chi^2$ based on the deviations from the nominal values of 
$\Delta{E}$, $m_{D}$, and $\Delta{m}$ if applicable.
 
\begin{figure}[t]
  \begin{center}
    \includegraphics[keepaspectratio=true,height=40mm]{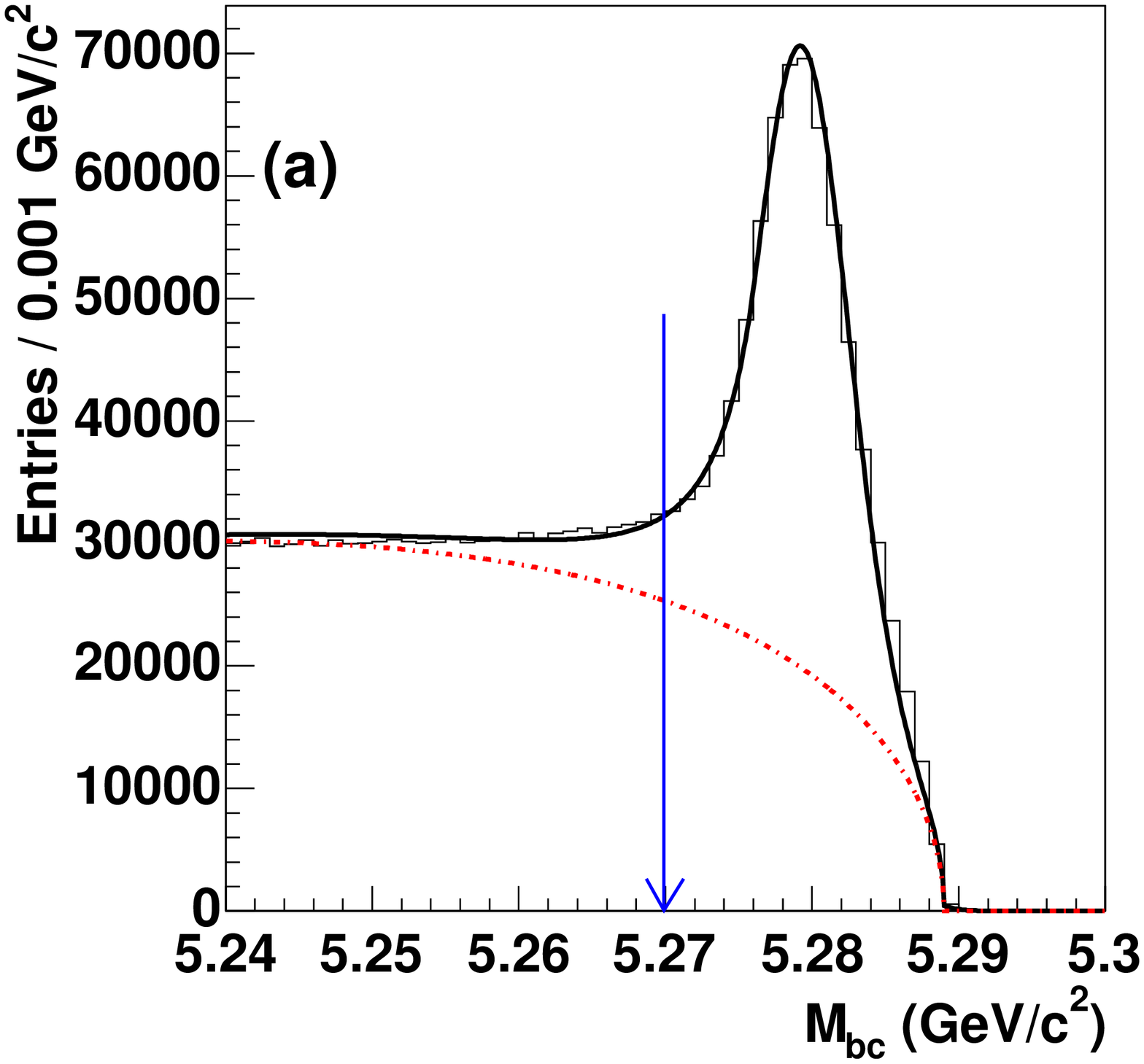}
    \includegraphics[keepaspectratio=true,height=40mm]{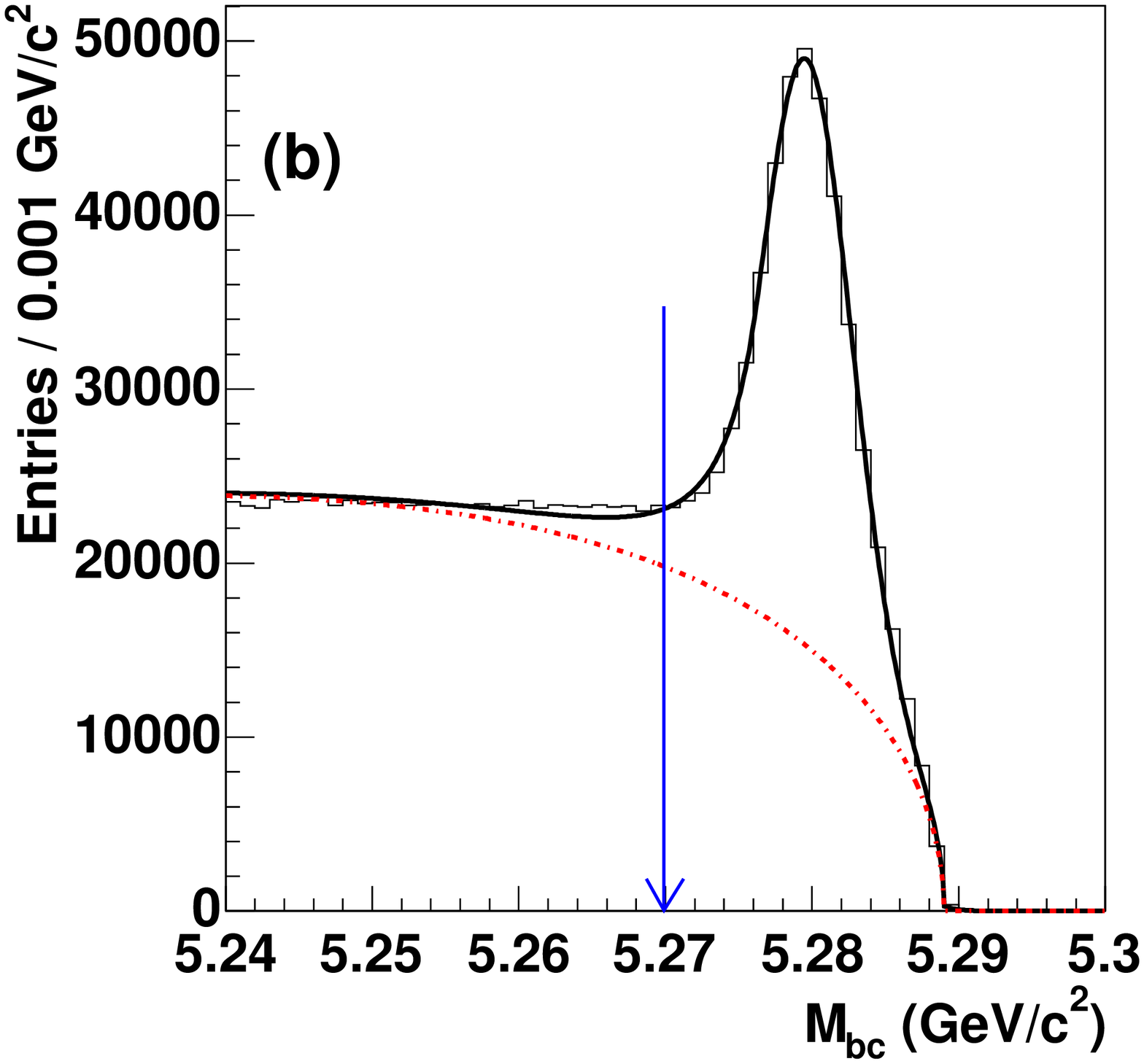}
  \end{center}
  \caption{$M_{\rm bc}$ distributions for fully reconstructed hadronic (a) $B^+$ and 
  (b) $B^0$ decays. The solid curve is the sum of the fitted signal and 
  background components, the dashed curve is the fitted background component. 
  The signal region is indicated by solid arrows.}
  \label{fig:fullrecon}
\end{figure}

Figure~\ref{fig:fullrecon} shows the distribution in $M_{\rm bc}$ of 
$B^+$ and $B^0$ candidates in the  $\Delta{E}$ signal region. The number of 
tagged events is obtained by fitting the distribution with empirical functions:
a Crystal Ball function for signal~\cite{Crystal} and 
an ARGUS function for background~\cite{Argus}. 
The fits yield $(4.26 \pm 0.17) \times 10^5$ $B^+$ , with a purity
of 55\%, and $(2.72 \pm 0.11) \times 10^5$ $B^0$, with a purity
of 50\%. These yields include cross-feeds between $B^+$ and $B^0$; 
these peak in $M_{\rm bc}$ and are not well separated by fitting.
From the MC simulation, 
we estimate the fraction of $B^0$ ($B^+$) events in the reconstruction of 
$B^+$ ($B^0$) to be 0.095 (0.090).  
The corrected yields of tags are:
$N_{\rm tag} = (3.88 \pm 0.20) \times 10^5$ for $B^+$ and 
$N_{\rm tag} = (2.46 \pm 0.12) \times 10^5$ for $B^0$. 

On the signal side, we reconstruct the following modes:
$B^ - \to D^{(*)0} \ell^{-} \bar{\nu}_{\ell}$, $D^{(*)+} \pi^- \ell^{-} \bar{\nu}_{\ell}$, 
and $\bar{B^0} \to D^{(*)+} \ell^{-} \bar{\nu}_{\ell}$, $D^{(*)0} \pi^+ \ell^{-} \bar{\nu}_{\ell}$. 
$\bar{B} \to D^{(*)} \ell^{-} \bar{\nu}_{\ell}$ decays are reconstructed 
as control samples to determine experimental resolutions, evaluate 
systematic errors, and normalize the results. 
All modes are reconstructed with all remaining particles after the 
full reconstruction of the tag. Flavor combinations of the two $B$ mesons
are restricted to $B^+B^-$, $B^0\bar{B^0}$, $B^0B^0$ and $\bar{B^0}\bar{B^0}$. 
We do not require opposite flavor for neutral $B$ mesons because of 
the effect of mixing. 

The selection of $D^{(*)}$ mesons is performed in the same manner as for 
the tag. We require that lepton candidates be identified as electrons or 
muons and have momentum greater than 0.6 GeV/$c$ 
in the laboratory frame. The electron and muon with largest momentum are 
selected as lepton candidates in each event.
The pion candidates are selected with a loose criterion, $P_{K/\pi} < 0.9$. 
We select the $B$ meson candidate formed with the best 
$D^{(*)}$ candidate (smallest $\chi^2$ of 
the deviations from the nominal values of $m_D$, and $\Delta{m}_{D}$ if applicable), 
and require that no charged 
track be left unused. Additionally, in the case of 
$\bar{B}^0 \to D^0 \pi^+ \ell^{-} \bar{\nu}_{\ell}$ 
and $D^{*0}({\to}D^0\pi^0, D^0\gamma) \pi^{+} \ell^{-} \bar{\nu}_{\ell}$ reconstructions, 
we require $M(D^0\pi^+) > 2.1~{\rm GeV}/c^2$ to veto events from 
$\bar{B}^0 \to D^{*+} \ell^{-} \bar{\nu}_{\ell}$.

\begin{figure}[t]
  \begin{center}
   \includegraphics[keepaspectratio=true,height=40mm]{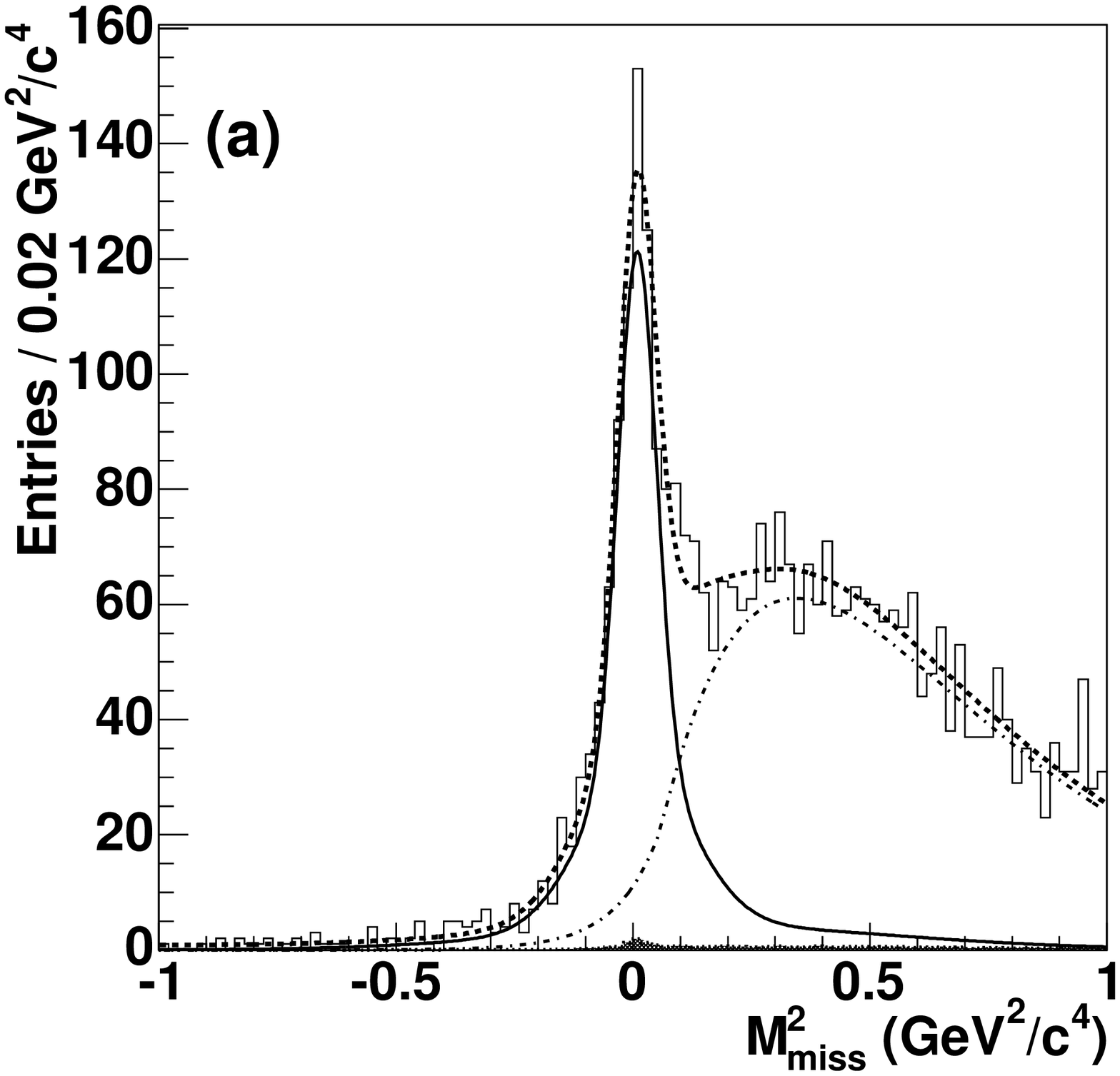}
   \includegraphics[keepaspectratio=true,height=40mm]{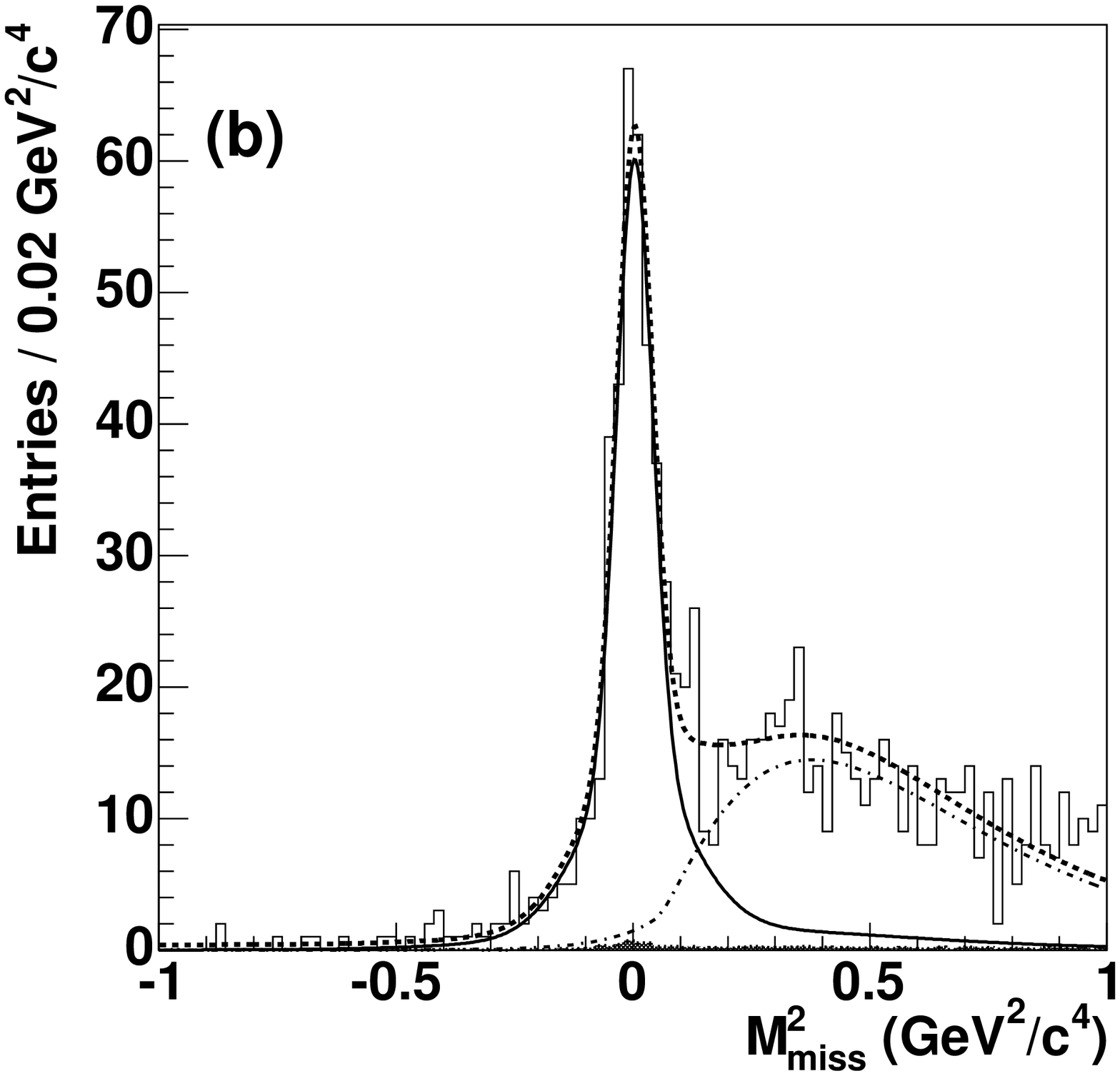}
   \includegraphics[keepaspectratio=true,height=40mm]{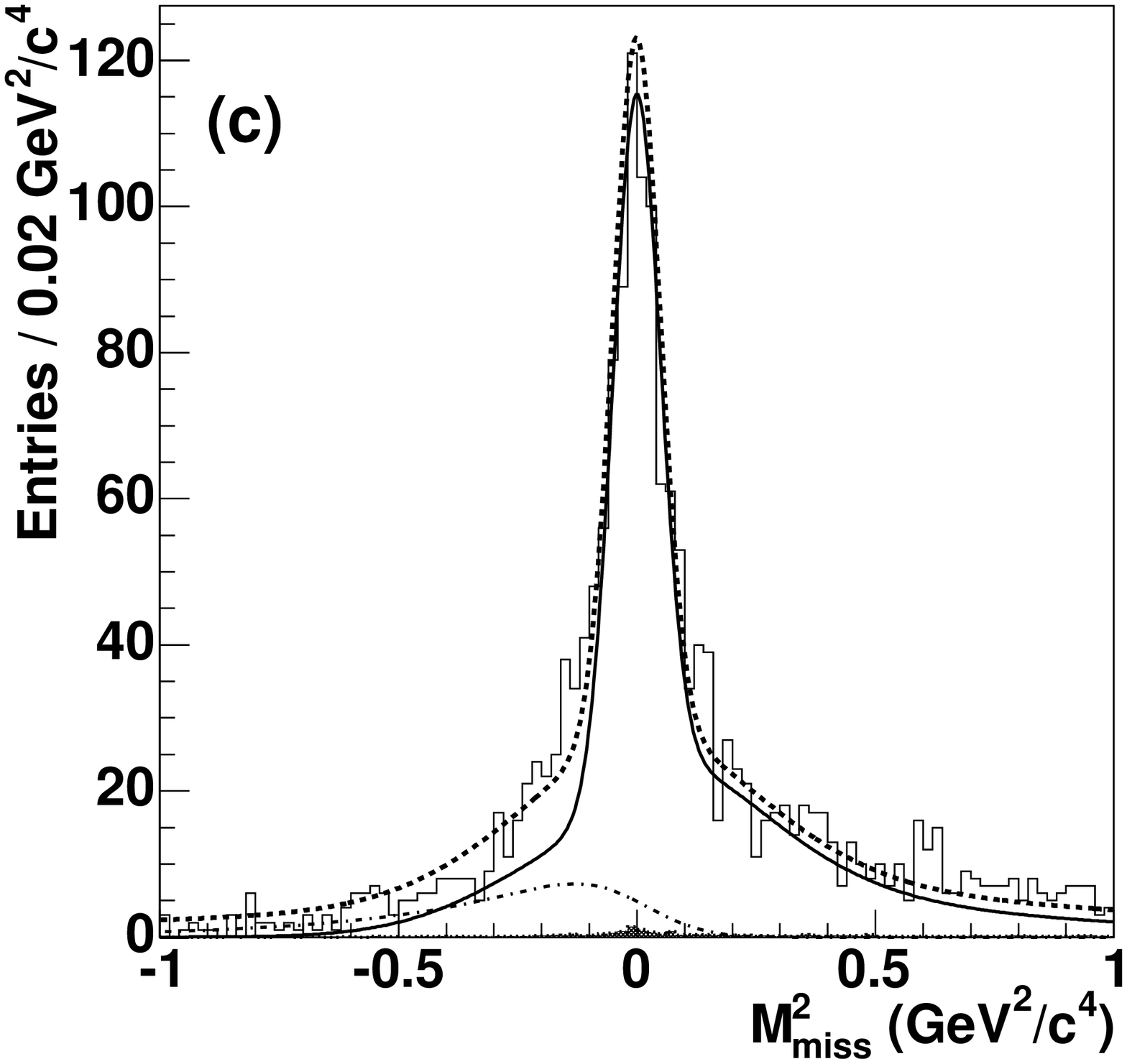}
   \includegraphics[keepaspectratio=true,height=40mm]{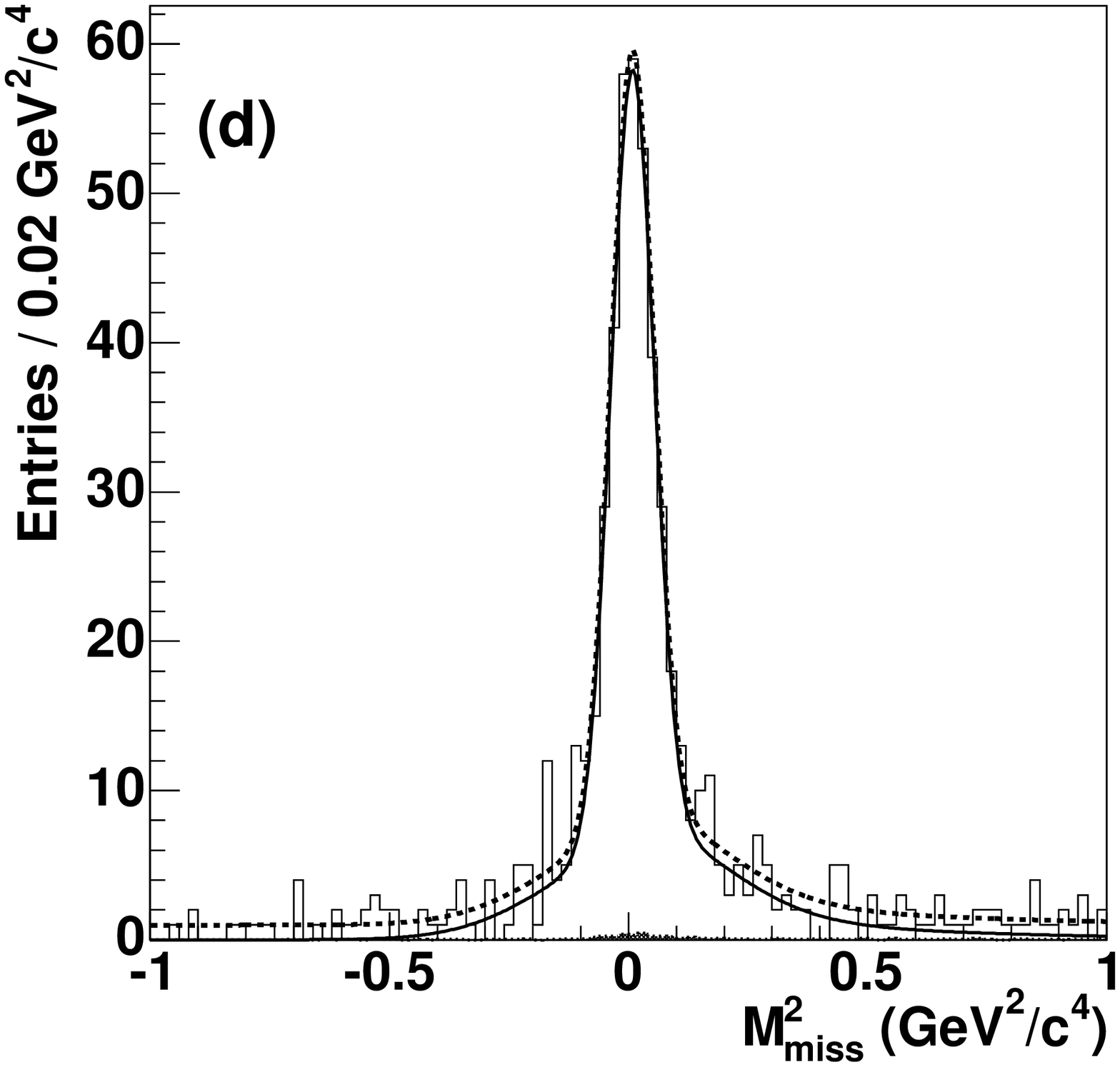}
  \end{center}
  \caption{ $M_{\rm miss}^2$ distributions for 
  (a) $B^{-} \to D^{0} \ell^{-} \bar{\nu}_{\ell}$, 
  (b) $\bar{B}^{0} \to D^{+} \ell^{-} \bar{\nu}_{\ell}$, 
  (c) $B^{-} \to D^{*0} \ell^{-} \bar{\nu}_{\ell}$ and 
  (d) $\bar{B}^{0} \to D^{*+} \ell^{-} \bar{\nu}_{\ell}$.
  Data are plotted as a histogram, and the fit result is overlaid as a dotted line. 
  Signal (solid curve), other semileptonic decays (dash-dotted) and 
  misidentified hadronic events (shaded histogram) are also shown. }
  \label{fig:miss2_dlnu}
\end{figure}

\begin{figure}[h]
  \begin{center}
   \includegraphics[keepaspectratio=true,height=40mm]{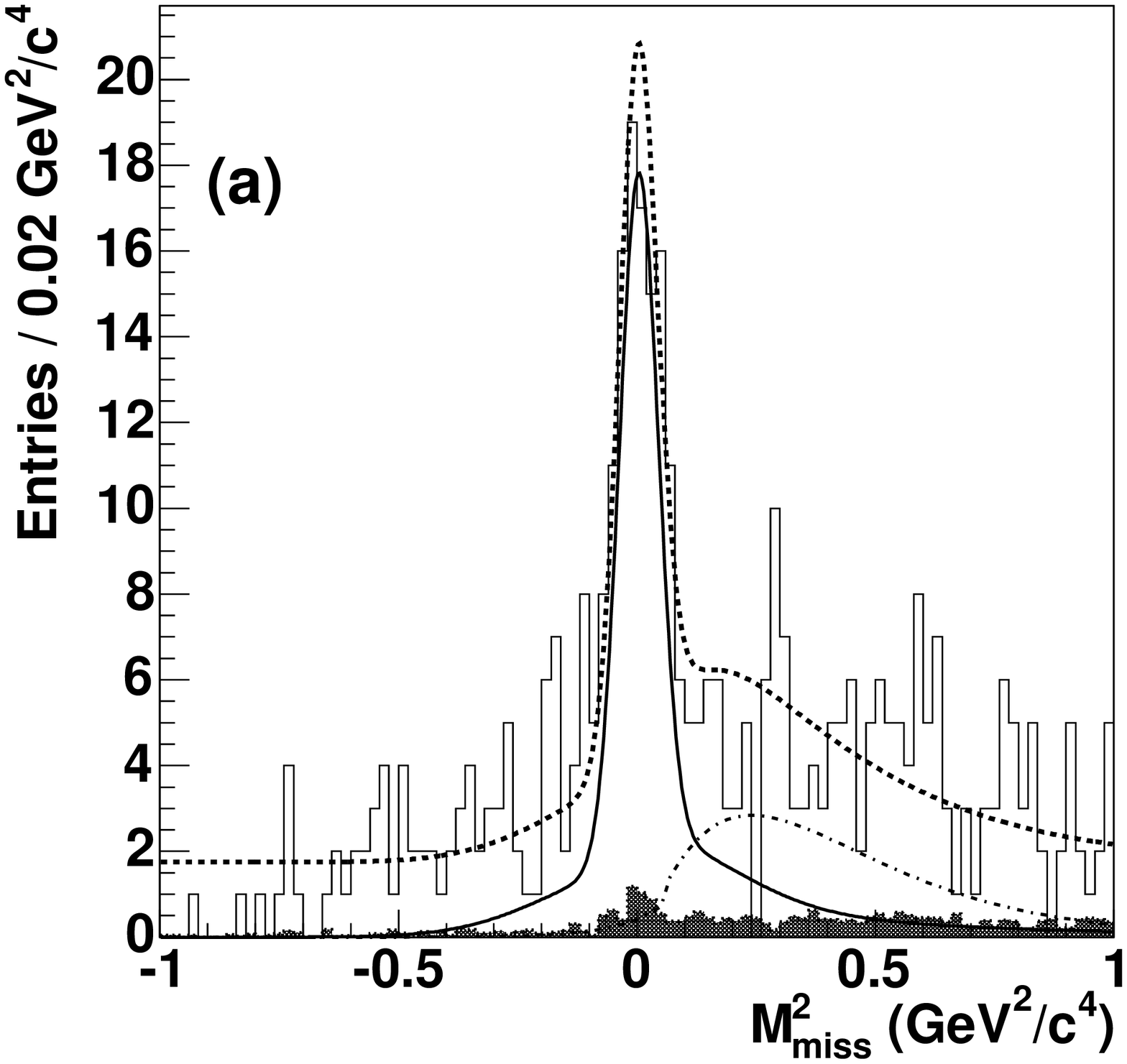}
   \includegraphics[keepaspectratio=true,height=40mm]{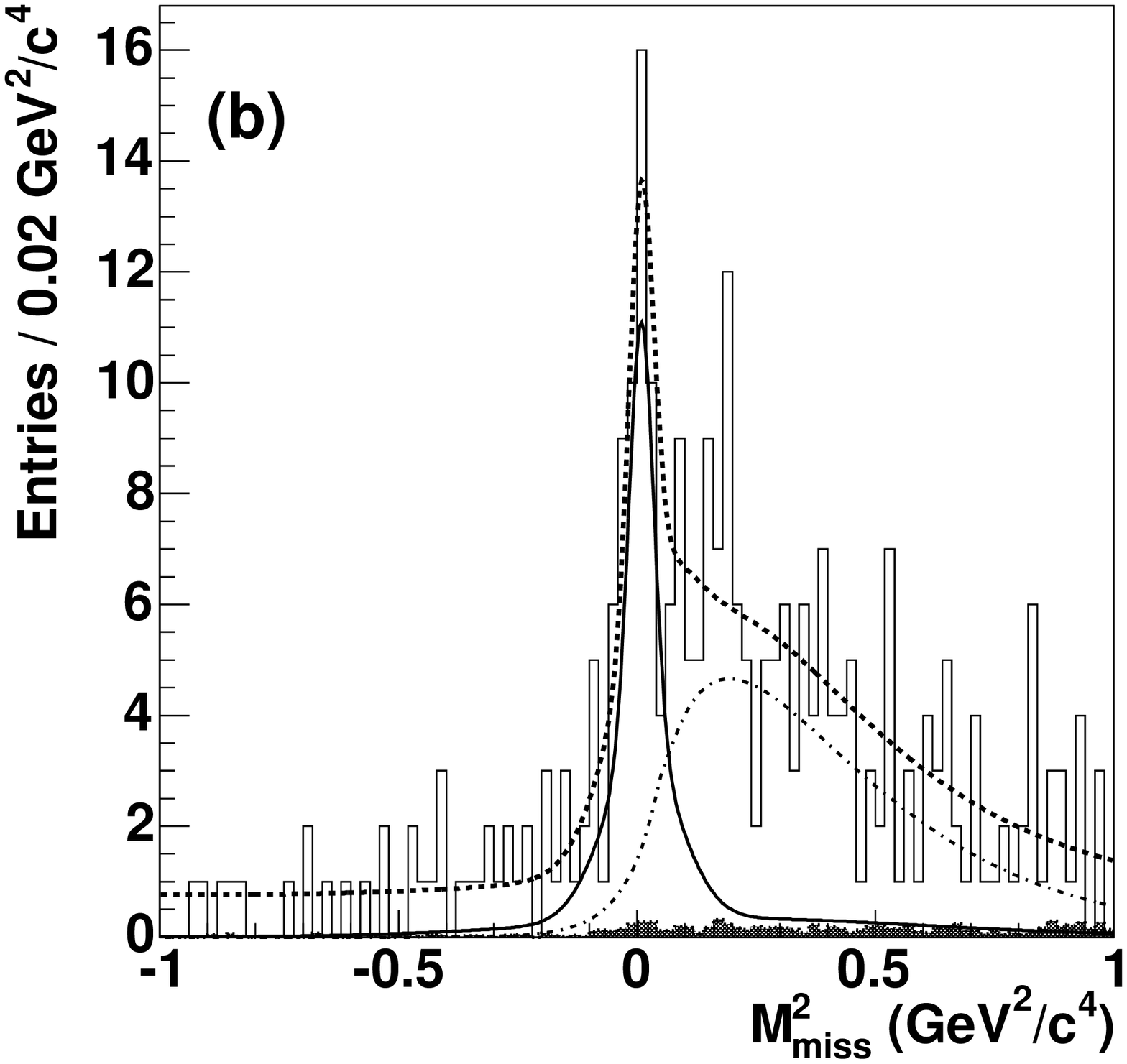}
   \includegraphics[keepaspectratio=true,height=40mm]{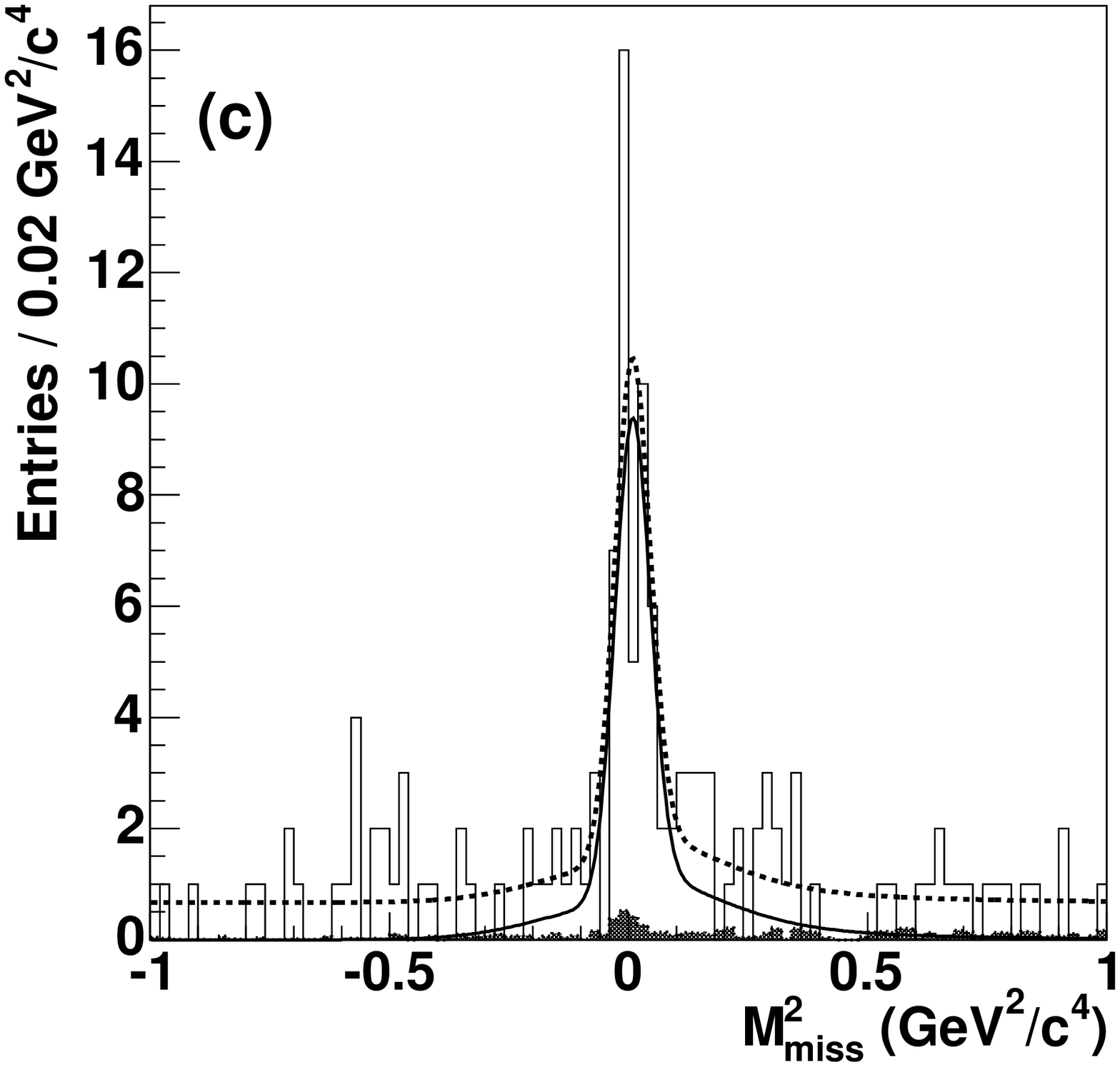}
   \includegraphics[keepaspectratio=true,height=40mm]{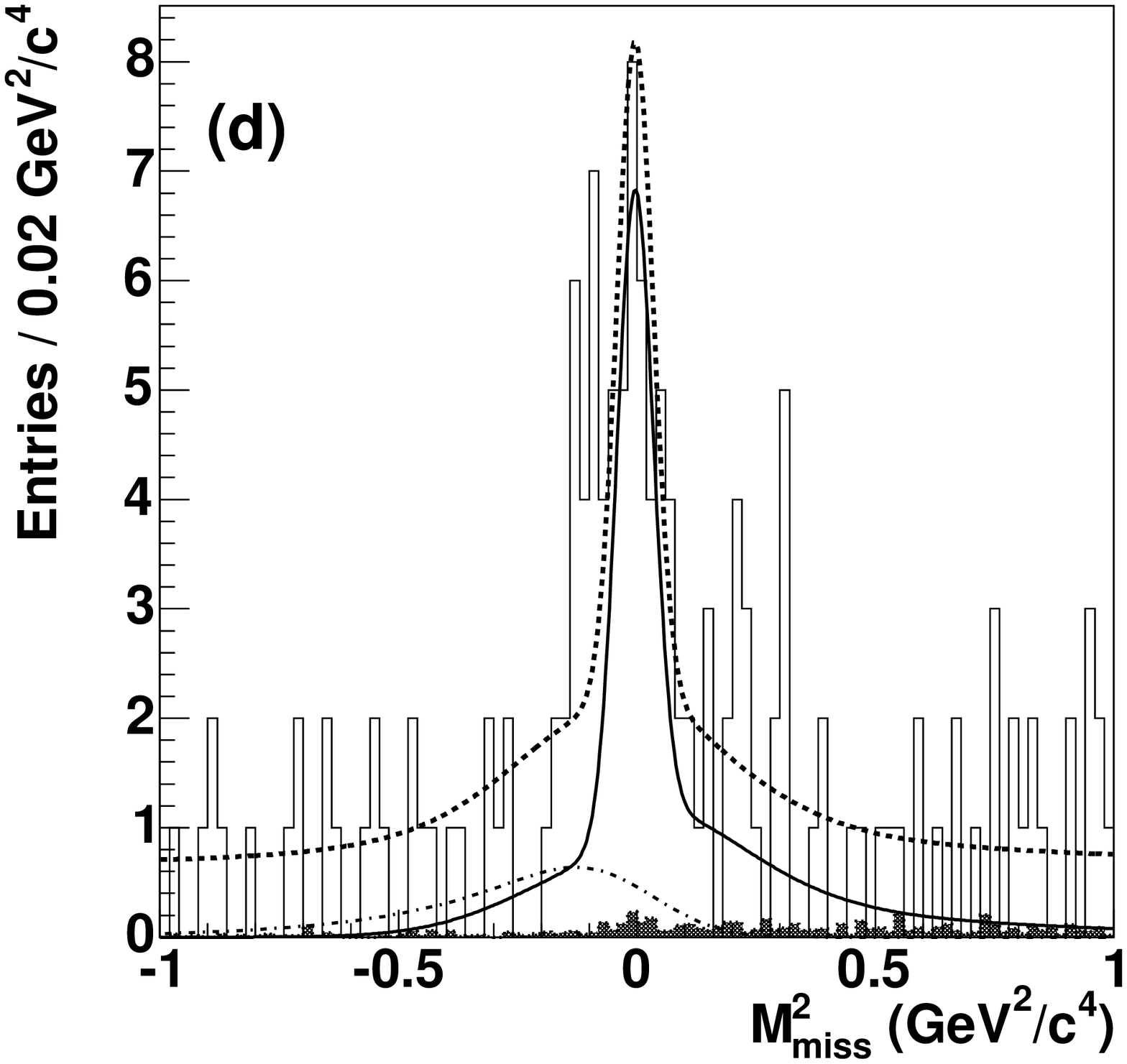}
  \end{center}
  \caption{ $M_{\rm miss}^2$ distributions for 
  (a) $B^{-} \to D^{+} \pi^{-} \ell^{-} \bar{\nu}_{\ell}$, 
  (b) $\bar{B}^{0} \to D^{0} \pi^{+} \ell^{-} \bar{\nu}_{\ell}$, 
  (c) $B^{-} \to D^{*+} \pi^{-} \ell^{-} \bar{\nu}_{\ell}$ and 
  (d) $\bar{B}^{0} \to D^{*0} \pi^{+} \ell^{-} \bar{\nu}_{\ell}$. 
  Distributions are shown as described in Fig.~\ref{fig:miss2_dlnu}. }
  \label{fig:miss2_dpilnu}
\end{figure}

The semileptonic decay is identified by the missing mass squared,
$M_{\rm miss}^2 =  ( E_{\rm beam} - E_{D^{(*)}(\pi)} - E_{\ell^{-}} )^2 - 
                   ( {\bf P}_{B_{\rm tag}} + {\bf P}_{D^{(*)}(\pi)} + {\bf P}_{\ell^{-}} )^2$, 
where $E_{\rm beam}$ is the beam energy, $E_{D^{(*)}(\pi)}$ and $E_{\ell^{-}}$ are
the $D^{(*)}(\pi)$ and lepton energies, and ${\bf P}_{D^{(*)}(\pi)}$ and ${\bf P}_{\ell^{-}}$
are the corresponding 3-momenta. ${\bf P}_{B_{\rm tag}}$ is the 3-momentum of 
the tag. All these variables are calculated in the $\Upsilon(4S)$ rest frame. 
For the signal, $M_{\rm miss}^2$ peaks around zero.
The resolution on $M_{\rm miss}^2$ ranges from $0.03$ to $0.07~{\rm GeV}^2/c^4$, 
depending on the mode. In contrast, previous analyses 
(e.g. by the ARGUS collaboration~\cite{Argus_Mmiss2}),
in which the ${\bf P}_{B_{\rm tag}}$ term was neglected, led to a resolution
of ${\sim}0.5~{\rm GeV}^2/c^4$ on $M_{\rm miss}^2$. 
This very good $M^2_{\rm miss}$ resolution allows reduction of the 
combinatorial background and to separate semileptonic decays that differ  
 by only one pion or $\gamma$ in the final state.
The distributions in $M_{\rm miss}^2$ for $\bar{B} \to D^{(*)} \ell^{-} \bar{\nu}_{\ell}$ 
and $D^{(*)} \pi \ell^{-} \bar{\nu}_{\ell}$ are shown in 
Figs.~\ref{fig:miss2_dlnu} and \ref{fig:miss2_dpilnu}. 
The signal events are clearly evident. 


We have investigated the backgrounds from $B\bar B$ and continuum events using 
MC simulation and off-resonance data. There are four main sources of
background:
{
\begin{enumerate}
\setlength{\itemsep}{0pt}\setlength{\parsep}{0pt}\setlength{\partopsep}{0pt}
\setlength{\parskip}{0pt}\setlength{\parindent}{10pt}
\item
Semileptonic $B$ decays where a pion or a photon is
missed (e.g. $\bar{B} \to D^{*} (\pi) \ell^{-} \bar{\nu}_{\ell}$ 
reconstructed as $\bar{B} \to D (\pi) \ell^{-} \bar{\nu}_{\ell}$ if
the soft $\pi^0$ or $\gamma$ from $D^{*}$ is missed). 
These distribute at high values of $M_{\rm miss}^2$ and can therefore be
distinguished from signal. 
\item
Semileptonic $B$ decays where a random photon candidate is used 
in $D^{*0}$ reconstruction 
(i.e. $\bar{B} \to D^{0} (\pi) \ell^{-} \bar{\nu}_{\ell}$ is reconstructed 
as $\bar{B} \to D^{*0} (\pi) \ell^{-} \bar{\nu}_{\ell}$). 
These events have a
lower value of $M_{\rm miss}^2$. 
\item 
Hadronic decays $\bar{B} \to D^{(*)} n h$ 
where one of the hadrons is misidentified as a lepton.
These events peak near $M_{\rm miss}^2=0$  and
therefore need special care as described below. 
\item
Random combinations.
As the tag is
fully reconstructed, this background is expected to be small. The distribution
in $M_{\rm miss}^2$ is studied using events in the $\Delta{E}$ side-band
($0.1~{\rm GeV} < \Delta{E} < 0.3~{\rm GeV}$) and off-resonance data. 
It is found to be consistent with a flat distribution in 
$M_{\rm miss}^2$.
\end{enumerate}
}
The signal yields are obtained by a binned maximum likelihood fit to 
the $M_{\rm miss}^2$ distributions in the interval $[-1.0,1.0]$~${\rm GeV}^2/c^4$.
The total fit function is:
$F(x\equiv M_{\rm miss}^2) = ( N_{\rm sig} + N_{\rm bkg3} ) S(x) + N_{\rm bkg1} B_1(x) + 
N_{\rm bkg2} B_2(x) + N_{\rm bkg4} B_4(x)$, where $N_{\rm sig}$ is the number
of signal events and the $N_{\rm bkg(n)}$ are the number of background events 
corresponding to categories 1--4 listed above. 
The signal shape $S(x)$ is a sum of two Gaussians and a single-sided 
exponential convolved with a Gaussian;
the smeared exponential is necessary to describe the upper 
tail, which is due to energy loss, mainly by radiation from the electron. 
$B_{i}(x)$ are normalized background shapes. 
For $B_1(x)$, we use a threshold function ($\propto x^{\alpha} \exp(-\beta x)$), 
plus an exponential function to describe the lower
tail component  at  $M_{\rm miss}^2 \simeq 0~{\rm GeV}^2/c^4$. 
$B_2(x)$ is a smeared exponential function. 
$B_4(x)$ is constant in the fitted interval.

$N_{\rm sig}$, $N_{\rm bkg1}$ and $N_{\rm bkg4}$ are floated in the fit,
while $N_{\rm bkg2}$ and $N_{\rm bkg3}$ are determined from data and
fixed. (According to our background categorization, 
$N_{\rm bkg1}$ ($N_{\rm bkg2}$) is non-zero only for 
$\bar{B} \to D (\pi) \ell^{-} \bar{\nu}_{\ell}$ 
($\bar{B} \to D^{*0} (\pi) \ell^{-} \bar{\nu}_{\ell}$) decays.) 
The shape parameters of $B_1(x)$ and $B_2(x)$ are fixed using 
MC simulations. The mean and width of $S(x)$ are floated 
in the fit to $\bar{B} \to D^{(*)} \ell^- \bar{\nu}_{\ell}$, and the 
observed discrepancies with MC simulations are used to fix the mean
and width of $S(x)$ for $\bar{B} \to D^{(*)} \pi \ell^{-} \bar{\nu}_{\ell}$.
Other shape parameters of $S(x)$ are determined from MC
simulations.

To determine $N_{\rm bkg2}$ we multiply the measured number of 
$\bar{B} \to D^{0} (\pi) \ell^{-} \bar{\nu}_{\ell}$ events by the probability
of such events faking $\bar{B} \to D^{*0} (\pi) \ell^{-} \bar{\nu}_{\ell}$, as
determined in MC simulations. 
To determine the contribution from misidentified hadronic decays ($N_{\rm bkg3}$),
we use a sample of events which are reconstructed as signal but where the lepton 
candidates do not pass the identification requirement.
We weight these events by lepton mis-identification rates binned in laboratory
momentum, as determined from a control sample of 
$D^{*+}{\to}D^0({\to}K^- \pi^+) \pi^+$ events. 
(In the case of $\bar{B}^0 \to D^{(*)+} \pi^- \ell^{-} \bar{\nu}_{\ell}$, we also 
consider the possibility that the prompt pion is misidentified as a lepton.)
$N_{\rm bkg3}$ is then determined by a fit to the $M_{\rm miss}^2$ distributions
of this special sample.

The results for $\bar{B} \to D^{(*)} \ell^{-}\bar{\nu}_{\ell}$ and
$\bar{B} \to D^{(*)} \pi \ell^{-}\bar{\nu}_{\ell}$ are shown in 
Fig.~\ref{fig:miss2_dlnu}, \ref{fig:miss2_dpilnu} and listed in 
Table~\ref{table:results_dlnu}, \ref{table:results_dpilnu}.
The statistical significance of the signal yield is defined as 
$\Sigma = \sqrt{2\ln(-L_{0}/L_{\rm max})}$, where $L_{\rm max}$ and $L_0$ denote 
the maximum likelihood value and likelihood value obtained assuming zero signal events, 
respectively. 
Signals are observed in all $\bar{B} \to D^{(*)} \pi \ell^{-}\bar{\nu}_{\ell}$ modes 
with a statistical significance of more than 7.

\begin{table}[htbp]
  \begin{center}
  \begin{tabular}{|l|c|c|c|c|c|}
    \hline
     Mode  &   $N_{\rm sig}$     & $N_{\rm bkg3}$ & $\epsilon$(\%)
                 & ${\cal{B}}$($10^{-2}$) & PDG ${\cal{B}}$($10^{-2}$) \\
    \hline\hline
  $D^0 \ell^{-} \bar{\nu}_{\ell}$ &  
  $1049 \pm 44$  & $9 \pm 4$ & 5.70 &  $2.37 \pm 0.10$   & $2.15 \pm 0.22$ \\
  $D^{*0} \ell^{-} \bar{\nu}_{\ell}$  & 
  $1419 \pm 59$  & $13 \pm 4$ & 3.02 & $6.06 \pm 0.25$    & $6.5 \pm 0.5$   \\
  $D^+ \ell^{-} \bar{\nu}_{\ell}$  &  
  $467 \pm 26$   & $4 \pm 2$ & 4.15 & $2.14 \pm 0.12$   & $2.14 \pm 0.22$  \\
  $D^{*+} \ell^{-} \bar{\nu}_{\ell}$ &  
  $476 \pm 25$   & $3 \pm 2$ & 2.05 & $4.70 \pm 0.24$   & $5.44 \pm 0.23$  \\
    \hline
  \end{tabular}
 \end{center}
 \caption{Signal yields, number of misidentified hadronic events, 
 efficiencies, and raw branching fractions 
 for $\bar{B} \to D^{(*)} \ell^{-} \bar{\nu}_{\ell}$, 
 compared with their world average values. 
 Systematic errors are not included.  }
 \label{table:results_dlnu}
\end{table}

\begin{table}[htbp]
  \begin{center}
  \begin{tabular}{|l|c|c|c|c|}
    \hline
  Mode  &  $N_{\rm sig}$  &  $N_{\rm bkg3}$ & $\epsilon$(\%) & $\Sigma$  \\
    \hline\hline
  $D^+ \pi^- \ell^{-} \bar{\nu}_{\ell}$        &  
  $142.1 \pm 16.5$ & $5.7 \pm 3.2$ & 3.41 & 12.4  \\
  $D^{*+} \pi^- \ell^{-} \bar{\nu}_{\ell}$     & 
  $62.5 \pm 9.7$   & $2.6 \pm 1.8$ & 1.38 & 9.9  \\
  $D^0 \pi^+ \ell^{-} \bar{\nu}_{\ell}$    & 
  $72.0 \pm 12.4$  & $1.1 \pm 1.7$ & 4.06 & 8.2   \\
  $D^{*0} \pi^+ \ell^{-} \bar{\nu}_{\ell}$ &  
  $62.7 \pm 11.6$  & $1.7 \pm 1.5$ & 2.09 & 7.2  \\
    \hline
  \end{tabular}
 \end{center}
 \caption{Signal yields, number of misidentified hadronic events, 
 efficiencies, and statistical significance 
 for $\bar{B} \to D^{(*)} \pi \ell^{-} \bar{\nu}_{\ell}$. }
 \label{table:results_dpilnu}
\end{table}

The raw branching fractions are calculated as ${\cal{B}} = N_{\rm sig}/(2\epsilon N_{\rm tag})$, 
where $N_{\rm sig}$ is the measured number of 
$\bar{B} \to D^{(*)}(\pi) \ell^{-}\bar{\nu}_{\ell}$ events and $N_{\rm tag}$ is 
the number of selected tags. 
The efficiency $\epsilon$ is defined as $\epsilon \equiv  \epsilon^{\rm signal} 
\times \epsilon_{\rm frec}^{\rm signal}/ \epsilon_{\rm frec}^{\rm generic}$, 
where $\epsilon_{\rm frec}^{\rm generic}$ is fully reconstructed 
tagging efficiency and $\epsilon_{\rm frec}^{\rm signal}$ is 
its efficiency in the case of one $B$ decays into the signal mode. 
The factor $\epsilon_{\rm frec}^{\rm signal}/ \epsilon_{\rm frec}^{\rm generic}$ 
accounts for the difference of $B$ tagging efficiency between the 
signal decay modes and generic $B$ decays and is estimated to be around 1.05
 depending on the mode. 
 The obtained raw branching fractions  in normalization modes, 
$\bar{B} \to D^{(*)}\ell^{-}\bar{\nu}_{\ell}$ are shown in Table \ref{table:results_dlnu}. 
These agree with the world average values quoted in Ref.~\cite{PDG2004}.

To calculate the branching fractions for $\bar{B} \to D^{(*)} \pi \ell^{-} \bar{\nu}_{\ell}$, 
we first find the ratio $R$ of the raw branching fractions for each 
of these decays to that of the $\bar{B} \to D^{(*)} \ell^{-} \bar{\nu}_{\ell}$  
decay mode with the same $D^{(*)}$ charge, then multiply $R$ by the world average values of 
${\cal{B}}(\bar{B} \to D^{(*)} \ell^{-} \bar{\nu}_{\ell})$~\cite{PDG2004}. 
The results are shown in Table \ref{table:results_dpilnu2}.

Systematic errors in the measurement of the branching fractions for 
$\bar{B} \to D^{(*)} \pi \ell^{-} \bar{\nu}_{\ell}$ are associated with 
the uncertainties in the signal yields, tag yields, and reconstruction 
efficiencies. Most of the systematic errors related to the reconstruction 
of the $D^{(*)}$ and the lepton, as well as the branching fraction of 
the $D^{(*)}$ decays, cancel out in the ratio $R$.

Systematic errors in the signal yields are estimated from  uncertainties in the
signal shape, background shape and number of misidentified hadronic events. 
For the signal shape,  
the mean and width in $\bar{B} \to D^{(*)} \pi \ell^{-} \bar{\nu}_{\ell}$ 
are shifted by their respective errors obtained from the 
control sample, and determined to be $2{-}4\%$, depending on the mode. 
For the background shape, we estimated the uncertainty by using a different shape.
For example, the main background components in $\bar{B} \to D \pi \ell^{-} \bar{\nu}_{\ell}$ 
are changed from $\bar{B} \to D^{*} \pi \ell^{-} \bar{\nu}_{\ell}$ to
$\bar{B} \to D^{**}( \to D^{*} \pi) \ell^{-} \bar{\nu}_{\ell}$. 
The uncertainty is $12\%$ for $\bar{B^0} \to D^{0}\pi^+ \ell^{-}\bar{\nu}_{\ell}$ 
due to larger background events, while it is smaller than $4\%$ 
for other decay modes.  For the number of misidentified hadronic events, 
we estimated the uncertainty to be $2{-}3\%$ 
by varying $N_{\rm bkg3}$ by its error.

The systematic error due to the uncertainty in the number of tags
and amount of flavor cross-feed have to be considered as well, since the branching 
fraction calculation involves ratios of different $B$~meson flavors. The former is
estimated to be $4\%$ by varying the background shapes used in the $M_{\rm bc}$ fit,
and the latter is estimated to be $3\%$ by varying the fraction of cross-feed by
its error.

The effect of uncertainty on reconstruction efficiencies mostly cancels since we take 
the ratio to a sample that differs from the signal  by only one pion.
We, therefore, assign a total of $1\%$ error due to tracking efficiency (based on a study
of partially reconstructed $D^{*}$ decays) and particle identification (based on a study 
of kinematically selected $D^{*+}\to{}D^{0}(\to{}K^{-}\pi^{+})\pi^{+}$ decays).
The uncertainty due to finite MC statistics used to model 
the signal is $2{-}3\%$. 

To estimate the uncertainty in efficiency due to the modeling in the 
MC simulation, we compared $\bar{B} \to D^{**}( \to D^{(*)} \pi) \ell^{-} \bar{\nu}_{\ell}$ 
and $\bar{B} \to D^{(*)} \pi \ell^{-} \bar{\nu}_{\ell}$.
The difference of 10\% was assigned as the error due to this effect.

The total uncertainty is the quadratic sum of all above contributions, and amounts
to $13{-}20\%$, depending on the mode. In the measurement of the absolute 
branching fractions, we quote an additional $4{-}10\%$ systematic error due to the error
on branching fractions of the normalization modes 
$\bar{B} \to D^{(*)} \ell^{-} \bar{\nu}_{\ell}$.

As a cross-check an independent analysis was performed with tighter requirements 
on $D^{(*)}$ reconstruction, $\Delta{}E$, and $M_{\rm bc}$ of 
the tag and a slightly different fitting procedure 
and efficiency calculation. Results of both analyses are consistent.

\begin{table}[htbp]
  \begin{center}
  \begin{tabular}{|l|c|c|}
    \hline
     Mode  &  $R$  & ${\cal{B}}$($10^{-2}$)  \\
    \hline\hline
  $D^+ \pi^- \ell^{-} \bar{\nu}_{\ell}$ & 
  $0.25 \pm 0.03 \pm 0.03$ & $0.54 \pm 0.07 \pm 0.07 \pm 0.06$ \\
  $D^{*+} \pi^- \ell^{-} \bar{\nu}_{\ell}$ &  
  $0.12 \pm 0.02 \pm 0.02$ & $0.67 \pm 0.11 \pm 0.09 \pm 0.03$ \\
  $D^0 \pi^+ \ell^{-} \bar{\nu}_{\ell}$    &  
  $0.15 \pm 0.03 \pm 0.03$ & $0.33 \pm 0.06 \pm 0.06 \pm 0.03$ \\
  $D^{*0} \pi^+ \ell^{-} \bar{\nu}_{\ell}$ &  
  $0.10 \pm 0.02 \pm 0.01$ & $0.65 \pm 0.12 \pm 0.08 \pm 0.05$ \\
    \hline
  \end{tabular}
 \end{center}
 \caption{Ratios and branching fractions for 
 $\bar{B} \to D^{(*)} \pi \ell^{-} \bar{\nu}_{\ell}$ decays. The first error is statistical, 
 the second is systematic. For ${\cal{B}}$, the third error is due to 
 the branching fraction uncertainties of $\bar{B} \to D^{(*)} \ell^{-} \bar{\nu}_{\ell}$.  }
 \label{table:results_dpilnu2}
\end{table}

We compute the total branching fractions of $\bar{B} \to D^{(*)} \pi \ell^{-} \bar{\nu}_{\ell}$
assuming isospin symmetry, 
${\cal{B}}(\bar{B} \to D^{(*)} \pi^0 \ell^{-} \bar{\nu}_{\ell} ) = 
\frac{1}{2} {\cal{B}}(\bar{B} \to D^{(*)} \pi^{\pm} \ell^{-} \bar{\nu}_{\ell} )$,
to estimate the branching fractions of $D^{(*)}\pi^0$ final states. 
We obtain
\begin{eqnarray*}
  {\cal{B}}_{D^{(*)} \pi \ell^{-} \bar{\nu}_{\ell}} (B^{-})
   &=& (1.81 \pm 0.20 \pm 0.20){\times}10^{-2}, \\
  {\cal{B}}_{D^{(*)} \pi \ell^{-} \bar{\nu}_{\ell}} (B^{0})
   &=& (1.47 \pm 0.20 \pm 0.17 ){\times}10^{-2},
\end{eqnarray*}
where the first error is statistical and the second is systematic.
Our measurements are consistent with the ALEPH result and significantly 
smaller than that of DELPHI. This clearly shows, as suggested by ALEPH's result, 
that the missing branching fraction in semileptonic $B$ decays is not fully covered 
by these excited states. Further searches must be carried out in 
$\bar{B} \to D^{(*)} \pi \pi \ell^{-} \bar{\nu}_{\ell}$ modes.

A study of the mass structure of $D^{(*)}\pi$ in 
$\bar{B} \to D^{(*)} \pi \ell^{-} \bar{\nu}_{\ell}$ 
decays is needed to understand the decay mechanics of 
$\bar{B} \to D^{**} \ell^{-} \bar{\nu}_{\ell}$, 
which would provide crucial tests of HQET and 
QCD sum rules~\cite{QCD_sumrule}.  
Belle has recently observed all four $D^{**}$ states 
($D_1^{\prime}$, $D_0^{*}$, $D_1$ and $D_2^{*}$) in 
the hadronic $B$ decay $\bar{B} \to D^{**} \pi$ \cite{dd_belle}. 
The corresponding semileptonic decay $\bar{B} \to D^{**} \ell^{-} \bar{\nu}_{\ell}$, 
however, has been observed only for 
$\bar{B^0} \to D_1^{0} ( \to D^{*+} \pi^{-} ) \ell^{-} \bar{\nu}_{\ell}$ 
by CLEO~\cite{CLEO2_ddlnu}. 
These contributions could be clarified by the method with 
fully reconstructed tags with a larger data set.

In conclusion, we have measured the branching fractions of 
$B^- \to D^{(*)+} \pi^- \ell^{-} \bar{\nu}_{\ell}$, and 
$\bar{B}^0 \to D^{(*)0} \pi^+ \ell^{-} \bar{\nu}_{\ell}$.
These decay modes have been clearly observed in a clean environment 
thanks to full reconstruction tagging, and the direct measurement 
of these branching fractions was achieved for the first time.

We thank the KEKB group for excellent operation of the
accelerator, the KEK cryogenics group for efficient solenoid
operations, and the KEK computer group and
the NII for valuable computing and Super-SINET network
support.  We acknowledge support from MEXT and JSPS (Japan);
ARC and DEST (Australia); NSFC (contract No.~10175071,
China); DST (India); the BK21 program of MOEHRD, and the
CHEP SRC and BR (grant No. R01-2005-000-10089-0) programs of
KOSEF (Korea); KBN (contract No.~2P03B 01324, Poland); MIST
(Russia); MHEST (Slovenia);  SNSF (Switzerland); NSC and MOE
(Taiwan); and DOE (USA).


\begin{thebibliography}{99}

\bibitem{PDG2004}
S.~Eidelman {\it et al.}, Physics Letters {\bf B592}, 1 (2004) \\
(Particle Data Group, http://pdg.lbl.gov/).

\bibitem{ALEPH_ddlnu}
D.~Skulls {\it et al.} (ALEPH Collab.),
Z. Phys. {\bf C73}, 601 (1997).

\bibitem{DELPHI_ddlnu}
P.~Abreu {\it et al.} (DELPHI Collab.),
Phys. Lett. {\bf B475}, 407 (2000).

\bibitem{Belle_Vxb}
Examples of these measurements can be found in
K.~Abe {\it et al.} (Belle Collab.),
Phys. Lett. {\bf B526}, 247 (2002) and
H.~Kakuno {\it et al.} (Belle Collab.),
Phys. Rev. Lett. {\bf 92}, 101801 (2003).

\bibitem{Belle}
A.~Abashian {\it et al.} (Belle Collab.),
Nucl. Instr. and Meth. A {\bf 479}, 117 (2002).

\bibitem{KEKB}
S.~Kurokawa and E.~Kikutani, 
Nucl. Instr. and Meth. A {\bf 499}, 1 (2003), and other papers included in
this volume.

\bibitem{Hadron_select} K.~Abe {\it et al.} (Belle Collab.),
Phys. Rev. {\bf D 66}, 032007 (2002).

\bibitem{GEANT}
R.~Brun {\it et al.}, GEANT3.21, CERN Report DD/EE/84-1 (1984).

\bibitem{Evtgen}
See the EvtGen package home page, \\
http://www.slac.stanford.edu/\~{}lange/EvtGen/.

\bibitem{HQET2}
J.~Duboscq {\it et al.} (CLEO Collab.), Phys. Rev. Lett. {\bf 76}, 3898 (1996).

\bibitem{Goity_Roberts}
L.~Goity and W.~Roberts, Phys. Rev. {\bf D 51}, 3459 (1995).

\bibitem{ISGW2}
D.~Scora and N.~Isgur, Phys. Rev. D {\bf 52}, 2783 (1995). 
See also N.~Isgur {\it et al.}, Phys. Rev. {\bf D 39}, 799 (1989).

\bibitem{R2}
C.~Fox and S.~Wolfram, Phys. Rev. Lett. {\bf 41}, 1581 (1978).

\bibitem{Argus}
H.~Alberecht {\it et al.} (ARGUS Collab.), Z. Phys. {\bf C 48}, 543 (1990).

\bibitem{Crystal}
J.~E.~Gaiser {\it et al.}, Phys. Rev. {\bf D 34}, 711 (1986).

\bibitem{Argus_Mmiss2}
H.~Alberecht {\it et al.} (ARGUS Collab.), Z. Phys. {\bf C 57}, 533 (1993).

\bibitem{QCD_sumrule}
A.~Le~Yaounanc {\it et al.},
Phys. Lett. {\bf 520}, 25 (2001).

\bibitem{dd_belle}
K.~Abe {\it et al.}, Phys. Rev. {\bf D69}. 112002 (2004).

\bibitem{CLEO2_ddlnu}
A.~Anastassov {\it et al.} (CLEO Collab.),
Phys. Rev. Lett. {\bf 80}, 4127 (1998).

\end{thebibliography}
\end{document}